%% file: main.tex
\documentclass[acmtog,screen=true,nonacm]{acmart}

\AtBeginDocument{%
  \providecommand\BibTeX{{%
    \normalfont B\kern-0.5em{\scshape i\kern-0.25em b}\kern-0.8em\TeX}}}

\setcopyright{acmlicensed}
\acmJournal{TOG}
\acmYear{2022}
\acmVolume{41}
\acmNumber{4}
\acmArticle{46}
\acmMonth{7}
\acmDOI{10.1145/3528223.3530173}

\acmSubmissionID{691}

\input{preamble}

\begin{document}

\input{sections/metadata_arxiv}

\input{figures/teaser.tex}

\maketitle

\input{sections/intro.tex}

\input{sections/related_work.tex}

\input{sections/method.tex}

\input{sections/results.tex}

\input{sections/conclusion.tex}

\begin{acks}
We would like to thank Romain Rouffet, Luc Chamerlat, Geoffrey Rosin, and Gaetan Lassagne for their time, suggestions, and helpful feedback.
\end{acks}

\bibliographystyle{ACM-Reference-Format}
\bibliography{macros,main}

\input{sections/appendix.tex}

\end{document}

%% file: preamble.tex
\usepackage{color}
\usepackage{enumitem} 
\usepackage{microtype} 
\usepackage{xspace}

\newcommand{\name}{\textsc{MatFormer}\xspace}
\newcommand{\oslot}{\texttt{out}}
\newcommand{\islot}{\texttt{in}}

%% file: sections/metadata_arxiv.tex
\title{\name: A Generative Model for Procedural Materials}

\author{Paul Guerrero}
\email{guerrero@adobe.com}
\orcid{0000-0002-7568-2849}


\author{Milo\v{s} Ha\v{s}an}
\email{mihasan@adobe.com}
\orcid{0000-0003-3808-6092}

\affiliation{
  \institution{Adobe Research}
  \country{}
}

\author{Kalyan Sunkavalli}
\email{sunkaval@adobe.com}
\orcid{0000-0002-6030-2348}

\author{Radom\'{i}r M\v{e}ch}
\email{rmech@adobe.com}
\orcid{0000-0002-5558-0327}


\author{Tamy Boubekeur}
\email{boubek@adobe.com}
\orcid{0000-0001-5985-0921}


\affiliation{
  \institution{Adobe Research}
  \country{}
}

\author{Niloy J. Mitra}
\email{n.mitra@cs.ucl.ac.uk}
\orcid{0000-0002-2597-0914}

\affiliation{
  \institution{Adobe Research}
  \country{}
}
\affiliation{
  \institution{University College London}
  \country{UK}
}

\renewcommand{\shortauthors}{Guerrero et al.}

\input{sections/abstract.tex}

\ccsdesc[500]{Computing methodologies~Image manipulation}
\ccsdesc[300]{Computing methodologies~Machine learning}

\keywords{node graphs, procedural materials, transformers, generative models}

%% file: sections/abstract.tex
\begin{abstract}

Procedural material graphs are a compact, parameteric,  and resolution-independent representation that are a popular choice for material authoring. 
However, designing procedural materials requires significant expertise and publicly accessible libraries contain only a few thousand such graphs.
We present \name, a generative model that can produce a diverse set of high-quality procedural materials with complex spatial patterns and appearance.  
While procedural materials can be modeled as directed (operation) graphs, they contain arbitrary numbers of heterogeneous nodes with unstructured, often long-range node connections, and functional constraints on node parameters and connections.
\name addresses these challenges with a multi-stage transformer-based model that sequentially generates nodes, node parameters, and edges, while ensuring the semantic validity of the graph.  
In addition to generation, \name can be used for the auto-completion and exploration of partial material graphs.
We qualitatively and quantitatively demonstrate that our method outperforms alternative approaches, in both generated graph and material quality.
\end{abstract}

%% file: figures/teaser.tex
\begin{teaserfigure}
  \includegraphics[width=\textwidth]{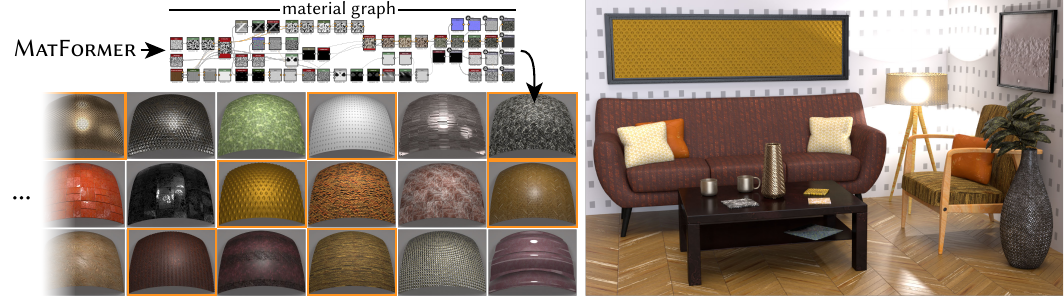}
  \caption{
  We present \name, a generative model for procedural materials that are represented as node graphs. \name  generates an arbitrary number of editable, resolution-independent materials (left), that can be used in realistic scenes (right). All materials in the scene were generated by our method; some of them are shown highlighted in orange on the left. Materials are generated as node graphs (top), where nodes correspond to image operators (each node shows the operator output), and edges control the flow of information between nodes. User can adjust parameters directly using the parameterized material graphs.}
  \label{fig:teaser}
\end{teaserfigure}

%% file: sections/intro.tex
\section{Introduction}

High-quality materials are an essential ingredient for creating virtual assets for a range of computer graphics applications including games, movies, and AR/VR. 
Procedural \textit{node graphs} are particularly popular as a controllable, resolution-independent material representation. 
By combining geometry with such procedural textures, artists regularly create realistic and compelling virtual assets. 

Procedural material design (e.g., using a product such as Adobe Substance 3D Designer) typically involves creating a directed node graph, referred to as a {\em material graph}.
Such graphs consist of a set of nodes---representing noise and pattern generators or operations on textures (e.g., filter kernels, transformations)---and edges---representing the flow of information from the output of the nodes to inputs of the subsequent nodes---finally producing image maps (e.g., roughness, normal, diffuse) for an analytic SVBRDF model. The output maps can be controlled by editing the parameters of the individual nodes.
With complex material definitions regularly needing 50+ nodes, authoring such graphs is a form of black magic, limited to a select handful of practitioners. Not surprisingly, the largest publicly-available texture dataset \cite{Substance_Share} has only a few thousand 
such definitions, and non-expert users mainly select from these limited options. Hence, there is a demand for automatically generating procedural materials, or assisting with their creation.

In the recent years, deep learning has pushed the state-of-the-art in generative models for images~\cite{Karras:2020:StyleGAN2}, animations~\cite{generativeAnimation:16}, videos~\cite{videoGen:19}, geometry~\cite{mo2019structurenet}; and even the direct generation of material maps has been explored~\cite{Guo:2020:MaterialGAN}. However, procedural materials, because of their controllability, are a desirable representation, and previous methods cannot be directly applied to generate them because of multiple challenges. 
First, unlike images/videos, material graphs do not have a regular structure and can have an arbitrary number of nodes with varying depth and breadth. 
Second, material graphs are typically composed of heterogeneous nodes that have different numbers and types of parameters, and different numbers of edge slots. 
Third, each node's input and output slots have different functional specifications, based on its (node) type, that need to be accounted for.
Lastly, material graphs contain long chains of operations, with long-range connections between distant nodes that are often critical to the appearance of the material.

In this work, we introduce \name, the \emph{first autoregressive generative model for material graphs} that addresses the challenges detailed above.
\name leverages a transformer-based~\cite{Vaswani:2017:Transformers} architecture to model a probability distribution over the space of procedural materials, and subsequently allows sample from this distribution.
We found the choice of transformers, as opposed to LSTM, GRU, or graph networks, to be particularly suitable in this context, as transformers effectively handle sparse long-distance connections between graph nodes. 
However, in order to model the specific structure of material graphs, \name does not generate them in a single pass. Instead, it runs in three sequential stages, each modeled with a dedicated transformer to capture dependencies: first, we generate a sequence of nodes; second, we generate parameters for each of the generated nodes; and finally, we generate directed edges connecting the input and output slots of the generated nodes.

\name is trained on a dataset of $2816$ procedural graphs \cite{Substance_Source} and can generate a diverse set of high-quality material graphs. In Figure~\ref{fig:teaser} we show example procedural materials \emph{automatically} generated from \name; please refer to the supplementary for a thousand such material graph generations. These materials exhibit a wide range of spatial patterns, material appearance,  and geometric detail. Moreover, each procedural material can be manipulated, using the node parameters, to generate many more variations. For example, in Figure~\ref{fig:teaser}~(right) a user chose $21$ materials from a palette of $200$ auto-generated graphs and combined them with 3D models to create photorealistic renderings of a complex scene. We demonstrate clearly, via various quantitative metrics, that \name outperforms alternate approaches in terms of the quality of the graphs it generates, as well as the plausibility and diversity of the materials that can then be sampled from these graphs. 

We also enable a novel auto-complete application where a user starts by picking a set of noise generators, and optionally a few associated starting nodes with connections, thus providing a partial material graph. Based on this specification, we can construct multiple completed graphs, that the user can iteratively explore and refine (please see Fig.~\ref{fig:autocomplete}). \name thus enables a fundamentally different form of material authoring that is considerably simpler and more intuitive that current procedural material design tools.

%% file: sections/related_work.tex
\section{Related Work}
\paragraph{Procedural materials}
Procedural modeling, given its relevance in synthesizing assets including patterns, shapes, buildings, landscapes, animations, materials, has a long history in computer graphics. 
Here, we focus on procedural methods only in the context of textures. These methods define functions that map spatial locations to pixel values \cite{Perlin:1985:Image,Peachey:1985:Solid,Worley:1996:Cellular,GLLD2012GNBE}. Other methods simulate complex physical processes like reaction-diffusion to generate textures \cite{Witkin:1991:Reaction}. 
All these methods have parameters that give users control over texture generation. 
Modern material authoring tools, like Adobe Substance Designer, expand on these ideas: they allow users to combine filter nodes, which represent simple image processing operations, to build graphs that process noise and patterns to generate complex, spatially-varying materials. However, designing such graphs requires significant time and expertise.

To address this, researchers have proposed \emph{inverse} procedural material design methods that fit the parameters of a procedural function to an exemplar texture image. Lefebrve and Poulin~\shortcite{Lefebvre:2000:Analysis} use heuristics based on image features to estimate the parameters of tile and wood procedural models. 
Motivated by texture synthesis approaches~\cite{GatysEB15}, researchers have also introduced neural version of material generators~\cite{henzler2020neuraltexture} to directly output texture/material images. 
More recently, Guo et al.~\shortcite{Guo:2020:Bayesian} use MCMC sampling to optimize for the parameters of procedural material models to fit a target photograph; however, their procedural are simple and hand-coded (PyTorch) programs. Hu et al.~\shortcite{Hu:2019:Inverse} propose training deep neural networks to predict the parameters of procedural node graphs given a captured image, and Hu et al.~\shortcite{hu:2022:inverse}  propose to reconstruct SVBRDF maps with a procedural representation that consists of procedural region masks combined with procedural noise functions inside each region.
However, this is not a generative model.
The recently proposed MATch framework~\cite{Shi:2020:Match} converts procedural node graphs into differentiable programs and uses stochastic gradient descent to fit the graph parameters to captured images. 

Methods like MATch assume that a procedural graph (or function) is given. In contrast, we present a \emph{generative} model that can \emph{create} new procedural material graphs from scratch. 
Because of the expertise required to manually create them, even the largest procedural material libraries \cite{Substance_Share,Substance_Source} have 1500-5000 assets. Instead, \name allows users to create plausible, custom graphs on the fly, dramatically expanding the space of material appearance that can be generated and enabling a new set of applications in material design, such as free generation and partial graph auto-complete.

\paragraph{Generative models in graphics} The recent years have seen an impressive advancement in the ability of generative models~\cite{Goodfellow:2014:GAN,Karras:2019:StyleGAN} to model 2D images. Furthermore, these generative models have been used in inverse pipelines, projecting images into their latent spaces \cite{abdal2019image2stylegan,zhu2020indomain,richardson2021psp}. Similarly, methods have been proposed for the generation of 3D shapes \cite{mo2019structurenet,Nash:2020:Polygen}. In the context of materials, several methods have been proposed \cite{li:2019:synthesizing,Guo:2020:MaterialGAN,zhou:2021:adversarial} to generate realistic per-pixel material maps that can be used for material capture by projecting a set of flash photographs of a material sample into the latent space. However, the results are image maps, limited in resolution, and not parameterized. In contrast, we develop on a generative model for directly outputting procedural material graphs, which can then be adjusted for further variations by adjusting its parameters.

Procedural node graphs can be represented as sequences of nodes with edges between them. This allows us to use auto-regressive models to generate them. In particular, we use the Transformer model \cite{Vaswani:2017:Transformers}, which has an  inbuilt self-attention mechanism, that has been applied to natural language processing applications. Such auto-regressive models have also been used to generate images \cite{Oord:2016:PixelRNN}, sketches~\cite{sampaio2000sketchformer}, geometry \cite{Nash:2020:Polygen}, and  layout~\cite{yang2021layouttransformer}. However, the structure of material graphs, with their arbitrary number of nodes, varying number of input and output edges, and functional constraints on these edges, makes material graph generation significantly more challenging than generating text, images, or meshes. As described later, we propose a three-pass generation algorithm that addresses these challenges.

\paragraph{Program and/or Graph Generation in Graphics}
In the context of generative models for irregularly structured domains, \cite{grass17,mo2019structurenet} proposed recursive neural networks for shape synthesis by modeling intra-shape dependencies in the form of adjacency, symmetry, and hierarchical relations, expressed as binary or, more generally, n-ary graphs.
Subsequently, these generators were extended to directly produce shape programs~\cite{jones2020shapeAssembly}, in hand-coded domain specific languages, using a GRU-based recurrent network. PolyGen~\cite{Nash:2020:Polygen} utilized transformers to capture long-range relations to directly output shapes in the form of triangle meshes. 
Fayolle and Pasko~\shortcite{fayolle2016csg} were one of the first to explore CSG program construction from point clouds based on genetic programming. Du et al.~\shortcite{inverseCSG_du18} map a CSG trees to a purely discrete representation that is suitable to apply state-of-the-art program synthesizers.
With access to large-scale CAD instructions (e.g., Fusion360~\cite{fusion360}), researchers have treated shapes as texts, and investigated usage of natural language generators to propose  DeepCAD~\cite{deepCAD21} for directly producing CAD models. 
Continuing this line of work we propose an auto-regressive generator that is specifically designed to tackle the unique challenges associated with handling material graphs.

%% file: sections/method.tex
\section{Method}

\begin{figure}
    \centering
    \includegraphics[width=\linewidth]{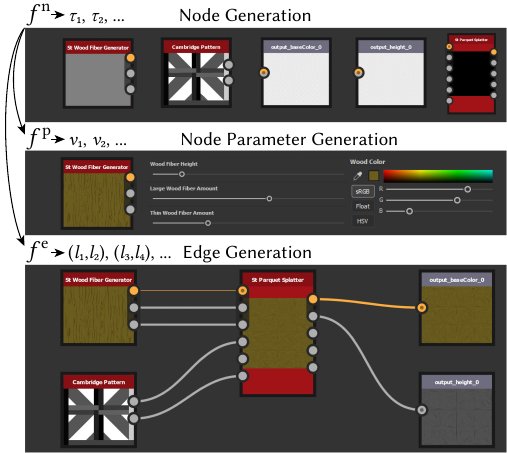}
    \caption{
   \textbf{ Material graph representation. }
    We generate a material graph in three steps, each step uses a Transformer-based generator $f$. First, we generate a set of operator types $\tau$ that describe the types of nodes in the graph. In the second step, we generate a sequence of node parameter values $v$ for each node, conditioned on the generated node types. Finally, we connect node pairs with edges based on a sequence of node index pairs $(l_i, l_{i+1})$ to construct the complete procedural material. Here, nodes are illustrated as boxes that display one of the node output images. The small circles to the left and right of any node denote input and output slots, respectively.} 
    \label{fig:overview}
\end{figure}

\name is a generative model for procedural materials; these materials are represented as directed node graphs that output a set of scalar- or vector image channels describing surface properties like albedo, roughness, metallicity, surface normals and height.

A node graph is a parametric model that generates an output given the parameters of each node. Nodes are instances of operations that define named \emph{input slots}, named \emph{output slots}, and a set of heterogeneous \emph{parameters}. Directed edges from output slots to input slots describe the flow of information in the node graph: an output of a given node is used as input by another node. Figure~\ref{fig:overview} shows an example of a node graph.
A node graph is a representation of a functional program, where functions with multiple input and output arguments are represented by nodes. The output arguments correspond to output slots of a node, while the input arguments are split into (a) parameters, which are constants, and (b) input slots, which are arguments that are outputs of other (preceding) functions. There are no loops in these programs. More details on node graphs are given in Section~\ref{sec:node_graphs}.

Generating node graphs for realistic materials poses a set of unique challenges, compared to generative models in more traditional pixel-based domains.
First, material graphs do not have a simple regular representation like an image grid and their node parameters are heterogeneous, i.e., each node type has a different set of parameters that may include scalars, vectors, and variable-length arrays.
Second, material graphs form long chains of operations that cause strong long-distance dependencies between nodes, making traditional approaches to sequence and graph generation ill-suited for the task, because information flow degrades with distance (e.g., Graph Neural Networks~\cite{scarselli2009gnn} and traditional Recurrent Neural Networks like GRUs~\cite{cho2014gru}).
Third, material graphs used in practice typically require a large number of nodes and node parameters.
Finally, material graphs have multiple functionally and semantically different input and output slots. Connecting the wrong slots may make the resulting graph not only sub-optimal, but invalid.

The above challenges motivate our design choices. Due to the long-distance dependencies between nodes, we choose Transformers~\cite{Vaswani:2017:Transformers} as our backbone network architecture, where information is available in equal fidelity from all parts of the graph, and an \emph{attention mechanism} is used to shape the information flow between different graph regions. Transformers originally come from natural language processing and operate on sequences, requiring a sequential ordering of the graph nodes.
Some previous work has tackled graph generation with Transformers in the context of 2D or 3D shapes~\cite{Nash:2020:Polygen, Para:2021:SketchGen}, by generating all nodes and their parameters as a single sequence.
This approach would be prohibitively expensive in our case, due to the memory and computation cost of the attention mechanism that is quadratic in the length of the sequence. Instead, we create node graphs in three sequence generation steps, corresponding to nodes, per-node parameters, and directed edges. Each step is implemented by a specialized Transformer, and the last two steps are conditioned on the output of the first step. Figure~\ref{fig:overview} illustrates this process, 

First, we generate nodes as a sequence of node types. However, multiple different node ordering strategies are possible, having different performance and applications (Section~\ref{sec:node_generation}).
Second, we generate the heterogeneous parameters of each node as separate sequences, conditioned on the node types that were generated in the previous step (Section~\ref{sec:param_generation}).
Finally,  we focus on edges. 
However, edges do not connect nodes directly, but rather connect input slots to output slots. A node can only have as many incoming edges as the number of input slots, given the node type; two nodes can be connected by multiple edges, as long as they terminate in different input slots. Inspired by recent work~\cite{Nash:2020:Polygen, Para:2021:SketchGen}, we generate edges using Transformers with Pointer Networks~\cite{Vinyals:2015:PointerNetworks}, which allow us to generate pointers into a predefined set of slots rather than nodes (Section~\ref{sec:edge_generation}). 

\subsection{Node Graphs}
\label{sec:node_graphs}

A node graph $g = (N, E)$ is a directed acyclic multigraph of image operators that consists of a set of nodes $N = \{n_1, n_2, \dots\}$ and edges $E=\{e_1, e_2, \dots\}$.
Given a set of node parameters, it outputs a set of material channels such as \emph{albedo} and \emph{roughness}. Figure~\ref{fig:overview} shows an example of a node graph.

Nodes $n = (\tau, P)$ are instances of image operators and are defined by an operator type $\tau$, and a set of parameters values $P$. The operator type $\tau$ is an index into a library of image operators $O = (o_1, o_2, \dots)$ that each take in a set of parameters and a set of images and output a set of images. The $k$-th operator is thus a function mapping input images into output images:
\begin{equation}
    (I^{\oslot}_1, I^{\oslot}_2, \dots) = o_k(I^{\islot}_1, I^{\islot}_2, \dots | P), \text{ with } P=(p_1, p_2, \dots),
\end{equation}
where each $I$ is a grayscale or RGB image and $p_j$ is a parameter that may be a scalar, vector, or variable-length array of vectors. The number of input images, output images, and the number and type of parameters are defined by the operator. 

A node $n_i$ of type $\tau_i$ has \emph{input slots} $(\islot^i_1, \islot^i_2, \dots)$ and \emph{output slots}, $(\oslot^i_1, \oslot^i_2, \dots)$, which are ports that edges can attach to, to provide input images and receive output images, respectively.
We call an operator that does not define any input images a \emph{generator}.

Directed edges $e=(\oslot^i_k, \islot^j_l)$ from output slots to input slots describe the flow of information in the node graph, i.e., the output image $I^{\oslot}_k$ of node $n_i$ is used as input $I^{\islot}_l$ of node $n_j$. Since an input slot can only accept a single input image, each input slot can only have one or zero incoming edges, while output slots can have an arbitrary number of outgoing edges.

Node graphs, which only define a partial ordering among the nodes, can be evaluated in any topological order. Note that cycles would cause infinite looping and are therefore not allowed. Further, not all input slots of a node need to be connected to edges. If no edge is attached, a default value is used for the corresponding operator input, typically an image of all zeros. The node outputs that are used for the final material channels are marked by connecting special \emph{output nodes} to their output slots. These output nodes do not perform any operation, they just annotate the graph output.

Next, we describe how to represent this graph with multiple token sequences, created with Transformer-based sequence generators.

\subsection{Transformers}
\label{sec:transformers}
We generate a node graph in three steps for nodes, node parameters, and edges, respectively. Transformers are used as generators in each step. Transformers~\cite{Vaswani:2017:Transformers} are sequence generators that were originally used for natural language processing. Nodes, node parameters, and edges are each generated as a sequence of discrete tokens $S=(s_1, \dots, s_m)$, one token $s_i$ at a time. More details on converting these graph components to sequence representations are given in the next sections.
Unlike earlier sequence generators like GRUs~\cite{cho2014gru} and LSTMs~\cite{hochreiter1997long}, Transformers use an attention mechanism that allows them to more accurately capture long-distance relations in a sequence.

\paragraph{Sequence generator.}
A \emph{Transformer-based generator} $f_\theta$ models the probability distribution over sequences $S$ as a product of conditional probabilities over individual tokens:
\begin{equation}
    p(S| \theta) := \prod_i p(s_i | s_{<i}, \theta),
\end{equation}
where $s_{<i} \coloneqq s_1, \dots, s_{i-1}$ denotes the partial sequence up to the token $s_{i-1}$. The model $f_\theta$ estimates a probability distribution over the possible value assignments for token $s_i$, conditioned on the partial sequence: $p(s_i | s_{<i}, \theta) = f_\theta(s_{<i})$. Once trained, we can directly sample the predicted distribution to obtain the token $s_i$. A complete sequence can then be generated, in a typical autoregressive setup, by repeatedly evaluating the model, starting from a special \emph{starting token} $\alpha$ and growing the partial sequence by one token in each step, until a special \emph{stopping token} $\omega$ is generated, or a maximum sequence length is reached.
We will apply some generators to multiple parallel sequences, denoted as $p^a_i, p^b_i = f_\theta(s^a_{<i}, s^b_{<i})$, where $p^*_i$ is short for $p(s^*_i | s^*_{<i}, \theta^*)$.

\paragraph{Semantic validity.}
Some value choices for a token may be semantically invalid. For example, a token that describes the end point of an edge should not refer to input slots that are already occupied by another edge. At inference time, we manually set generated probabilities $p(s_i | s_{<i}, \theta)$ for invalid choices to zero, and re-normalize the remaining probabilities. Validity checks for each step are described in Appendix~\ref{app:semantic_validity}.

\paragraph{Conditioning and sequence encoders.}
We condition the generation of node parameter sequences and edge sequences on the generated node sequence. Conditioning on a sequence requires a sequence encoder.
We use \emph{Transformer-based encoders} $g_\phi$, which have an architecture that is nearly the same as the generator described above, 
but each step takes as input the whole sequence and a token index and outputs an embedding of the token that may include information about the whole sequence: $\bar{s}_i := g_\phi(i, S)$, where $\bar{s}_i$ is a sequence-aware embedding of the token $s_i$.
All our generators $f_\theta$ and encoders $g_\phi$ are implemented as GPT-2 models~\cite{radford2019language}. To avoid notational clutter, we will omit the parameters $\theta$ and $\phi$ from the generator and encoder models $f$ and $g$ from here on.

\paragraph{Positional encoding.}
Both generators and encoders receive a \emph{positional encoding} of the tokens in a sequence as input. A positional encoding consists of additional sequences that are used as input, but do not need to be generated, since they can be derived from the other sequences. They are application-specific and provide additional context for each token in the sequence, such as the sequence index of a token.

\subsection{Node Generation}
\label{sec:node_generation}

In the first step, we generate the operator types $\tau$ of all nodes $n=(\tau, P)$ in a graph as sequence $S^\text{n} = (\alpha, \tau_1, \tau_2, \dots, \omega)$, using a Trans\-former-based model $f^\text{n}$. $\alpha$ and $\omega$ are the sequence starting and stopping tokens defined in Section~\ref{sec:transformers}.

\paragraph{Sequence representation.} A canonical ordering of the node sequence makes the generation task easier, as it provides more consistency between data samples, but also makes the model more inflexible, as it limits the diversity of the partial sequences $s_{<i}$ that the model is trained on. We experiment with four ordering strategies with different degrees of consistency; from most to least consistent:
\begin{itemize}
\small
    \item[$\pi_\text{r}$:] A back-to-front breadth-first traversal, starting from the output nodes, moving opposite to the edge directions from child nodes to parents, and traversing the parents in the order of a node's input slots. 
    \item[$\pi_\text{rr}$:] Reversed $\pi_\text{r}$ (i.e. from last to first node of $\pi_\text{r}$). 
    \item[$\pi_\text{b}$:] A front-to-back breadth-first traversal, where children of a node are visited in the order of the node's output slots. We randomize the order of children that are connected to the same output slot. 
    \item[$\pi_\text{t}$:] A random  but valid topological node ordering. 
\end{itemize}
We found that different orderings are suitable for different applications. We use $\pi_\text{r}$ for unconditional generation, since it is the most consistent, and $\pi_\text{rr}$ for our graph autocompletion, where a front-to-back ordering is beneficial.

\paragraph{Positional encoding and generation.} In addition to the sequence $S^\text{n}$, we define two sequences as positional encoding. The sequence $S^\text{ni} = (1, 2, 3, \dots)$ provides the global position of a token in the sequence and $S^\text{nd} = (0, d_1, d_2, \dots,  0)$ describes the depth of each node in the graph, where $d_i$ is the graph distance from $n_i$ to the closest output node when using $\pi_\text{r}$ or to the closest generator node for all other orderings. Since we cannot obtain the depth from the sequence $S^\text{n}$ alone, we estimate it as additional output sequence during generation.
The model $f^\text{n}$ uses all three sequences as partial input sequences and is trained to output probabilities for the next operator type and depth:
\begin{equation}
p^\text{n}_i, p^\text{nd}_i = f^\text{n}(s^\text{n}_{<i}, s^\text{nd}_{<i}, s^\text{in}_{<i}).
\end{equation}

\begin{figure}[t]
    \centering
    \includegraphics[width=\linewidth]{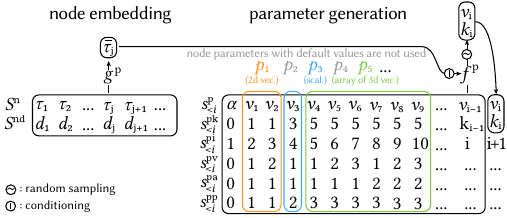}
    \caption{
    \textbf{Parameter generation.}
    The parameter list $P=(p_1, p_2, \dots)$ of a node $n_j$ is flattened into several sequences (right). Only parameters with non-default values are added to the sequence. Sequences are generated one token at a time by the Transformer-based generator $f^\text{p}$, which is conditioned on an embedding of $n_j$ that is computed by the sequence encoder $g^\text{p}$.}
    \label{fig:parameter_generation}
\end{figure}

\subsection{Parameter Generation}
\label{sec:param_generation}

Next, we generate the parameters $P$ of nodes $n=(\tau, P)$ as a sequence $S^\text{p} = (\alpha, v_1, v_2, \dots, \omega)$ of scalar values $v$, using the Trans\-former-based model $f^\text{p}$. One parameter sequence is defined per node in a graph, with parameters sorted alphabetically by name. Figure~\ref{fig:parameter_generation} illustrates the process. 

\paragraph{Conditioning on the node type.}
Parameter values depend on the operator type of the node $n_i$ they are generated for, and to a lesser extent on the other nodes in the graph. We condition generation of each sequence $S^\text{p}$ on a sequence-aware embedding $\bar{\tau}_j$ of the operator type $\tau_j$ that also includes information about the other nodes in the scene. The embedding is computed by a Transformer-based encoder $\bar{\tau}_j = g^\text{p}(j, S^\text{n}, S^\text{nd})$ that is trained jointly with the generator $f^\text{p}$.

\paragraph{Sequence representation.}
Typically, only a small fraction of node parameters are modified by an artist, the remaining parameters are left at their default values. We shorten the sequences $S^\text{p}$ significantly by only including parameters that are not at their defaults.
To identify which node parameter a given value $v_i$ in our shortened sequence refers to, we define a second sequence $S^\text{pk} = (0, k_1, k_2, \dots, 0)$ of indices $k$ into the full list of node parameters,
i.e. value $v_i$ corresponds to the node parameter with index $k_i$.

Node parameters are heterogeneous and may be scalars, vectors, and variable-length arrays of vectors. We flatten all parameters into a single sequence of scalars. Given the node type and the sequence $S^\text{pk}$, we can reconstruct parameters from the flattened sequence, since a node type and the parameter index fully define the type and vector dimension of a parameter. 
For variable length arrays of vectors, we obtain the array length by dividing the number of values generated for the parameter by the parameter's vector dimension. If the number is not evenly divisible, we discard the last values to make the number divisible.
Since transformers operate with discrete token values, continuous parameters are quantized uniformly between their minimum and maximum observed values in the dataset. In our experiments, we use $32$ quantization levels.

\paragraph{Positional encoding and generation.}
The positional encoding consists of the global token position sequence $S^\text{pi}$, and the sequences $S^\text{pv}$, $S^\text{pa}$, $S^\text{pp}$, which describe the index of the vector element, the index of the array element, and the index of the parameter associated with a token, respectively. Probabilities over parameter values and indices are thus generated as:
\begin{equation}
p^\text{p}_i, p^\text{pk}_i = f^\text{p}(s^\text{p}_{<i}, s^\text{pk}_{<i}, s^\text{pi}_{<i}, s^\text{pv}_{<i}, s^\text{pa}_{<i}, s^\text{pp}_{<i}\ |\ \bar{\tau}).
\end{equation}
Appendix~\ref{app:conditioning} gives architecture details of the conditional model $f^\text{p}$.

\subsection{Edge Generation}
\label{sec:edge_generation}
\begin{figure}[t]
    \centering
    \includegraphics[width=\linewidth]{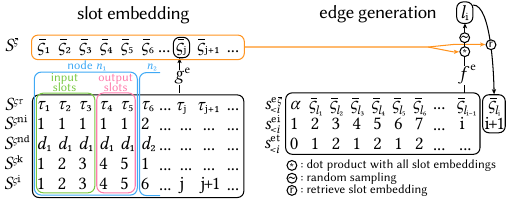}
    \caption{
    \textbf{Edge Generation. }
    Edges are generated by a Transformer-based generator $f^e$ as a sequence of index tuples that reference a sequence of input/output slots. These indices are generated using the mechanism proposed in Pointer Networks~\cite{Vinyals:2015:PointerNetworks}. Slots embeddings are computed using the sequence encoder $g^e$, based on a set of sequences that uniquely identify each slot.}
    \label{fig:edge_generation}
\end{figure}

In the last step, we generate all edges $e=(\oslot, \islot)$ in a graph as a sequence $S^e=(\alpha, l_1, l_2, l_3, l_4, \dots, \omega)$ of slot indices using the Transformer-based model $f^\text{e}$. Each pair of consecutive indices forms an edge. Figure~\ref{fig:edge_generation} illustrates edge generation.
\paragraph{Slot embedding.}
We create a sequence of all input and output slots in the graph. An edge is a pair of indices into this sequence. A straightforward approach to edge generation would be generate a sequence of slot indices directly using a standard Transformer generator. However, this would not give the generator any information about the slots in the graph, thus it would be hard for the transformer to reason about slot connectivity. Instead, we follow the work of~\cite{Vinyals:2015:PointerNetworks} and operate in the space of learnt slot embeddings.
We compute embeddings $S^{\bar{\varsigma}} = (\bar{\varsigma}_1, \bar{\varsigma}_2, \dots)$ for every slot using a Transformer-based encoder $g^\text{e}$ from multiple input sequences that together uniquely describe each slot in a node graph:
$S^{{\varsigma}\tau}$, $S^{{\varsigma}\text{ni}}$, $S^{\varsigma\text{nd}}$ are sequences of operator types, node indices, and node depths, respectively, and provide information about the node associated with each slot, while sequence $S^{\varsigma\text{k}}$ provides the index of the slot inside the node. A global token position sequence $S^{\varsigma\text{i}}$ provides additional positional encoding. The slot embedding is then computed as $\bar{\varsigma}_j = g^\text{e}(j, S^{\varsigma\tau}, S^{\varsigma\text{ni}}, S^{\varsigma\text{nd}}, S^{{\varsigma}\text{k}}, S^{{\varsigma}\text{i}})$. $g^\text{e}$ is trained jointly with $f^\text{e}$.

\paragraph{Pointer Network.}
To generate indices into a list of embeddings, we use the approach proposed in Pointer Networks~\cite{Vinyals:2015:PointerNetworks}. Our model $f^\text{e}$ outputs a feature vector $\bar{l}_i$ in each step instead of a probability distribution. This feature vector acts as a query into the list of slot embeddings $S^{\bar{\varsigma}}$. Affinities between the query and the embeddings are computed as dot products and normalized into a probability distribution: $p^e_i = \text{Softmax}_j(\bar{\varsigma}_j \cdot \bar{l}_i)$, where $\cdot$ denotes the dot product and $\text{Softmax}_j$ denotes the softmax over all indices $j$. This distribution over slots is sampled to obtain the slot index $l_i$. The partial sequence of slot embeddings selected in the previous steps $S^{\text{e}{\bar{\varsigma}}}$ is used as input to $f^\text{e}$, giving the model information about existing edges.

\paragraph{Positional encoding and generation.} The positional encoding for the edge sequence consists of the global token position sequence $S^{\text{ei}}$ and the tuple index of a token inside an edge $S^{\text{et}} = (0, 1,2,1,2,\dots, 0)$. Probabilities over slots indices are then generated as:
\begin{equation} 
p^e_i = \text{Softmax}_j\left( \bar{\varsigma}_j\ \cdot\ f^\text{e}(s^{\text{e}{\bar{\varsigma}}}_{<i}, s^{\text{ei}}_{<i}, s^{\text{et}}_{<i})\right).
\end{equation}

\subsection{Training Setup}
All models are trained with a binary cross-entropy loss over the probabilities estimated by the generators $g$. We use teacher forcing at training time, i.e. we use ground truth tokens for the partial sequences $s_{<i}$. Each step is trained separately using ground truth nodes as input for the node parameter and edge generation steps. We stop training once a minimum of the validation loss is reached to avoid overfitting. Additional details are provided in Appendix~\ref{app:arch_details}.

%% file: sections/results.tex
\section{Results}

We evaluate \name in two applications: \emph{unconstrained generation} of materials and \emph{guided authoring}, where our generative model assists an artist in material graph authoring by providing suggestions for possible completions of a partially constructed material graph. We first provide a few details about the dataset and the proceeded to experiments with the two applications.

\paragraph{Dataset.}
We use a subset of the Substance Source dataset~\cite{Substance_Source}
to train our models, which is made of ${\sim}4$k material graphs that were generated by experienced artists and are regularly used by a large community of professional artists. We filter out graphs with nodes that have a complex parameterization, such as nodes that are parameterized by a small pixel shader, leaving us with ${\sim}3.5$k graphs. From these graphs, we remove graphs with an exceedingly large number of nodes ($>400$) or edges ($>700$), or with nodes that have an unusually large number of input ($>21$) or output slots ($>14$), resulting in a dataset of ${\sim}2.8$k graphs.

\begin{figure*}
    \centering
    \includegraphics[width=\textwidth]{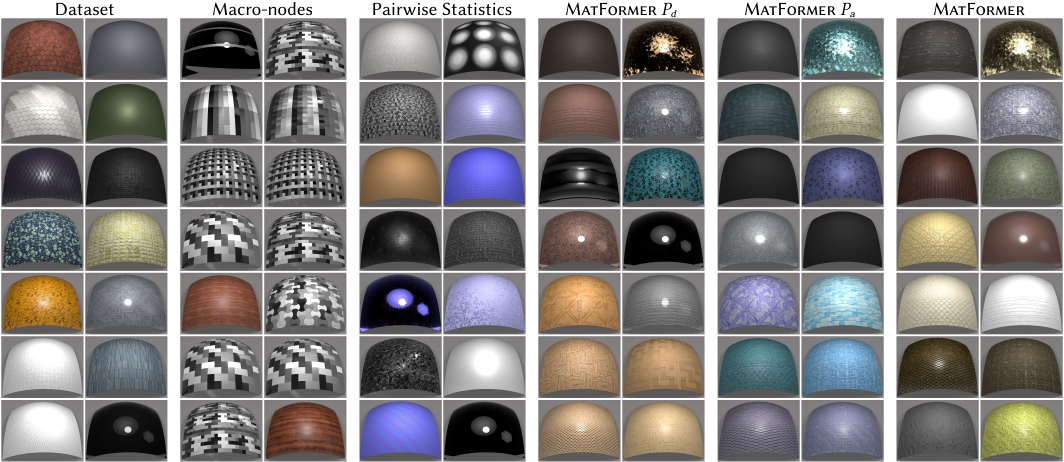}
    \caption{
    \textbf{Qualitative comparison.} 
    We compare 14 random samples taken from \name to random samples from the dataset, the baselines, and the parameter ablations. From left to right: (i)~dataset samples; (ii)~generation from single \emph{macro-nodes}; (iii)~generation using pairwise node connectivity statistics; (iv)~\name $P_d$, our method without the node parameter generation step and all parameters are at their defaults; (v)~\name $P_a$, our method with all parameters at the dataset average; and 
    (vi)~\name, our full method. Due to the long-range dependencies between nodes, our multistage Transformer-based approach generates material graphs that produce more diverse and realistic materials.}
    \label{fig:qualitative_comparison}
\end{figure*}

We augment the dataset with random node parameter variations. Specifically, we perturb each parameter value $v$ uniformly in the range $[0.8v, 1.2v]$.
We create $100$ graphs with random parameter perturbations for each graph in the original dataset, giving us ${\sim}280$k graphs. Since we are interested in generation, we only keep a small validation set of 500 graphs, generated by perturbing the parameters of 5 distinct graphs in the original dataset.
Random examples of augmented dataset graphs are shown in Figure~\ref{fig:qualitative_comparison}, left.

\subsection{Unconstrained Generation}

We demonstrate the ability of our method to generate a wide range of diverse and realistic materials by generating ${\sim}1$k samples (see supplementary material) and comparing them to several baselines, both qualitatively and quantitatively.

\subsubsection{Baselines.}
We are not aware of any existing complete generator for procedural materials; therefore we introduce our baselines.

First, the \emph{Macro-nodes} baseline uses graphs that consist only of a single \emph{macro node}. A macro node is a node that is defined by a subgraph and typically performs complex tasks that are needed frequently, and that would need to be re-created in many graphs. Macro nodes can be quite powerful, sometimes allowing for fairly complex output on their own. We compare to this baseline to show that the complexity seen in our graphs does not originate solely from the functionality encapsulated in these macro nodes.

Second, we compare to the \emph{Pairwise Statistics} baseline that uses pairwise node and slot connectivity statistics to generate a material graph. The baseline first collects dataset statistics that allowing us to create a probability distribution over the output slots and operator types that connect to each input slot on each operator type (including a \emph{null} type if the input slot is not connected). Starting from the output nodes, we can then iteratively grow the graph, going through each unconnected input slot in turn, and deciding which slot on which operator type it should connect to, if any. Once decided, we can either connect the input slot to a newly created node of the given operator type, or connect it to any existing node of the given type, as long as edges do not form a cycle. We repeat this process until all input slots have been visited, or until a maximum node count is reached. After the graph nodes and edges have been generated, we create material parameters using our parameter generator.

\subsubsection{Ablations.}
We evaluate the necessity for our parameter generator, and how different node orders influence the generation performance.

Two ablations compare our node parameter generation approach to simpler strategies.
The ablation \name $P_d$ sets all node parameters to their default values, while \name $P_a$ sets them to the average observed in our dataset, and adds the same random perturbation we use to augment our dataset.

Additionally, we evaluate each of the node ordering strategies ($\pi_\text{r}, \pi_\text{rr}, \pi_\text{b}, \pi_\text{t}$) proposed in Section~\ref{sec:node_generation}. We discuss the ablation results together with the baseline comparisons in Section~\ref{sec:discussion}.

\begin{figure*}
    \centering
    \includegraphics[width=\textwidth]{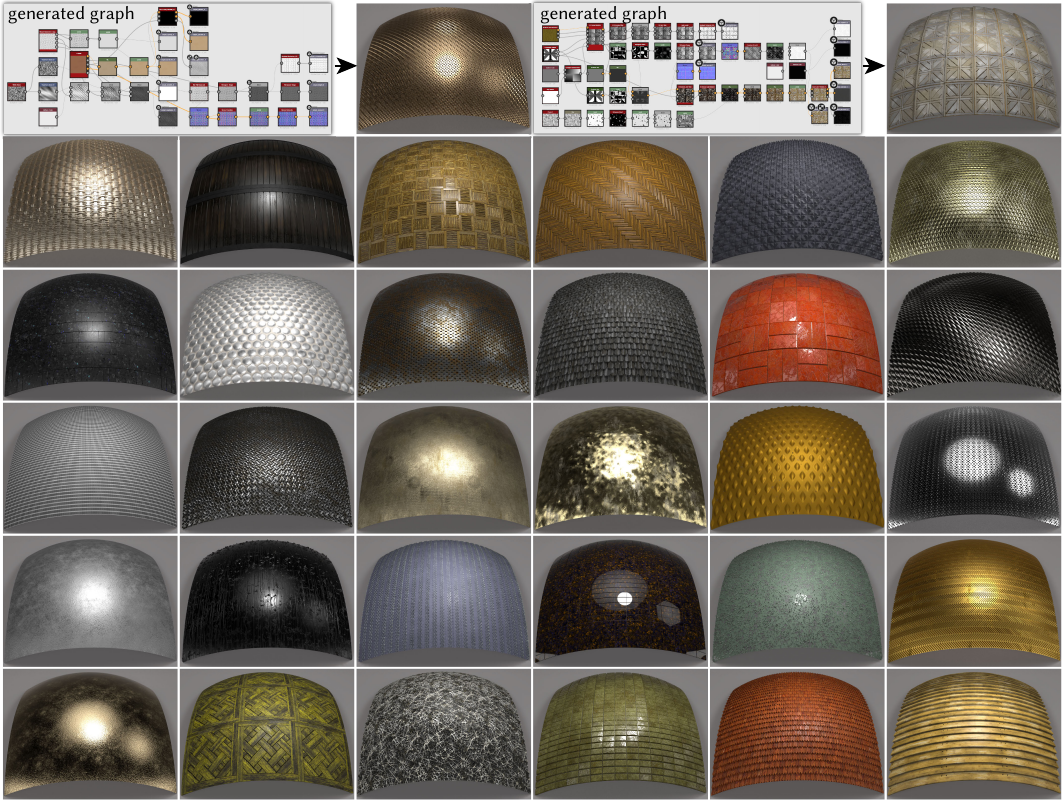}
    \caption{
    \textbf{Material graph generation. }
    We show several materials obtained from our automatically generated material graphs. In the top row we illustrate the material graphs of two examples. Boxes are nodes displaying the node output; input slots are on the left side, and output slots on the right side of each node. 
    Please note the wide variety of spatial patterns, normal and height variations (bumps), and specular highlights. 
    Please refer to the supplemental for 1k additional materials. 
    }
    \label{fig:qualitative_results}
\end{figure*}

\subsubsection{Metrics.}

We use four quantitative metrics.

(i) The \textit{graph statistics} $\mathcal{E}_{\text{g}}$ over the material graph structure, measuring various graph properties including the longest path and the typical graph distance between each pair of operator types. A full description is given in Appendix~\ref{app:graph_stats}. We represent each of the statistics with a histogram and compare the histograms of our generated graphs with those of the dataset using the Earth Mover's Distance (EMD)~\cite{rubner1998metric}; lower is better. 

(ii) The \emph{Frechet Inception Distance} (FID)~\cite{heusel2017gans} $\mathcal{E}_{\text{fid}}$ over the generated material samples. FID measures the similarity of two image distributions using simple statistics over the activations of an Inception network~\cite{szegedy2015going} when applied to an image set. We render material samples on a flat plane that is facing the camera, with a spot light illuminating the sample from the same direction as the camera, and compute the FID between our generated samples and the dataset distribution; lower is better. 

(iii) The \textit{normalized nearest neighbor style distance} $\mathcal{D}_{\text{nns}}$ computes the distance from each rendered material of a generated graph to the closest dataset sample, using a style distance~\cite{gatys2015neural} based on the Gram matrix of VGG~\cite{simonyan2014vgg} feature maps. We normalize this distance by dividing it with the average distance from each dataset sample to its closest neighbor in the dataset. Values significantly below $1.0$ indicate overfitting, while values significantly above $1.0$ indicate a distribution of generated samples that is not fully aligned with the dataset distribution, thus values around $1.0$ are ideal.

(iv) The \textit{normalized nearest neighbor edit distance} $\mathcal{D}_{\text{nne}}$ is analogous to $\mathcal{D}_{\text{nns}}$, but uses the edit distance between material graphs as distance metric (insertions and removals of both nodes and edges).
Due to the large computational cost of the graph edit distance, and since simple graphs are less likely to show interesting structural differences, we only consider graphs with $\ge 50$ nodes.

\begin{table}[b!]
  \centering
  \caption{
  {\bf Quantitative evaluation and comparison.} 
  We compare \name to two baselines, and evaluate its performance under different ablations. As metrics, we measure diversity and plausibility of the generations using FID scores ($\mathcal{E}_\text{fid}$); structure of the generated materials using graph statistics ($\mathcal{E}_\text{g}$); and novelty of the generations using a normalized distance to the nearest neighbors in the training dataset using both a style metric ($\mathcal{D}_\text{nns}$) and a graph edit distance ($\mathcal{D}_\text{nne}$), where values around $1.0$ are optimal. Please refer to the text for details. 
  }
  \begin{tabular}{r c c c c c}
    \toprule
  & $\mathcal{E}_{\text{g}}\downarrow$ & $\mathcal{E}_{\text{fid}}\downarrow$ & $\mathcal{D}_{\text{nns}}$ & $\mathcal{D}_{\text{nne}}$ \\ 
  \midrule
    \multicolumn{1}{r}{\textbf{Baselines}}\\
    Macro-nodes                  & 0.142 & 232.0 & 0.65 & n/a \\ 
    Pairwise Statistics          & 0.068 & 86.9 & 0.93 & 2.16\\ 
    \midrule
     \multicolumn{1}{r}{\textbf{Ablations}}\\
    \name $P_d$ (n. order $\pi_\text{r}$) & 0.047 & 94.2 & 0.88 & 1.52 \\ 
    \name $P_a$ (n. order $\pi_\text{r}$) & 0.047 & 76.4 & 1.04 & 1.52 \\ 
    \name (n. order $\pi_\text{t}$) & 0.068 & 66.4 & 0.93 & 2.21 \\ 
    \name (n. order $\pi_\text{b}$) & 0.070 & 68.4 & 0.84 & 2.08 \\ 
    \name (n. order $\pi_\text{rr}$) & 0.060 & 62.8 & 0.80 & 1.84 \\ 
    \midrule
    \name (n. order $\pi_\text{r}$) & \textbf{0.046} & \textbf{48.6} & \textbf{1.01} & \textbf{1.34} \\ 
    \bottomrule
  \end{tabular}
  \label{tbl:free_generation} 
\end{table}

\begin{figure*}[t]
    \centering
    \includegraphics[width=\textwidth]{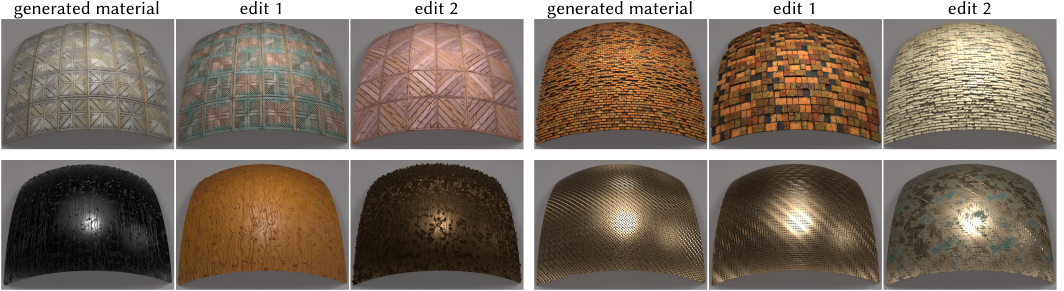}
    \caption{
    \textbf{Editing generated materials. }
    Our generated materials being procedural, they are, by construction, fully editable. We show two edits for each of four generated materials. Each edit was done in under $10$ minutes and no more than 10 node parameters were changed per edit.}
    \label{fig:editing}
\end{figure*}

\subsubsection{Evaluation and Discussion.}
\label{sec:discussion}
Quantitative comparisons to all baselines and ablations with ${\sim}1$k generated graphs are shown in Table~\ref{tbl:free_generation}, and qualitative comparisons are given in Figure~\ref{fig:qualitative_comparison}. For the quantitative comparison, we have trained four versions of our method, one for each node order $\pi_r$, $\pi_{rr}$, $\pi_b$, $\pi_t$ described in Section~\ref{sec:node_generation}, while for the qualitative comparison we use only $\pi_r$.

The results confirm that macro-nodes by themselves do not produce satisfactory results. Many macro-nodes produce output that needs further processing by additional nodes, such as the grayscale structures visible on many materials. Some materials, like the two wood materials, are reasonable but lack interesting structure and detail. The high FID scores and low distance $D_{nns}$ confirm the trend we see in the qualitative results. By construction, this baseline does not have graphs with $\ge50$ nodes, thus $D_{nne}$ cannot be computed.

The generator based on pairwise statistics performs better, but since it lacks the ability to model long-distance relationships, it can often get stuck with unrealistic outputs. The blue tint visible in many materials is likely due to normal conversion operators being applied incorrectly to a color channel. This is reflected in the high FID score. The high nearest neighbor distance $\mathcal{D}_\text{nne}$ shows that generated graphs are too far away from the manifold of dataset graphs.

The ablation \name $P_d$ with parameter values at their defaults uses node order $\pi_r$, so the most direct comparison is to the last row of Table~\ref{tbl:free_generation}. We can see that the graph structure metrics are comparable, since it is not affected by parameter generation. The FID score, however is significantly higher; we believe that this is due to the lack in diversity we can see in the qualitative results, resulting from parameters always being set to their default values. The FID score improves slightly in the ablation \name $P_a$, where parameter values are set to the average value observed in the dataset plus a perturbation, but the FID is still significantly higher than when using our parameter generator. The style nearest neighbor distance is higher here due to less realistic materials, as evidenced by the high FID score.

Among the four node orderings, the back-to-front ordering $\pi_r$ is the clear winner. We can see that there is a trend of increasing performance when going from most ambiguous ordering ($\pi_t$) to least ambiguous ($\pi_r$). Compared to the baselines, \name has a clear advantage in all the metrics.

Unless noted otherwise, we will use the model trained with node order $\pi_r$ for all unconstrained generation results, and the best-per\-forming front-to-back node order $\pi_{rr}$ for all user-guided authoring results.

Figure~\ref{fig:qualitative_results} shows several additional qualitative results of our method. We can see a diverse range of materials and structures, including metallic materials, bricks, wooden panels, etc. We show the corresponding generated material graphs for the two topmost examples. The supplementary material contains graph visualization and rendered materials for ${\sim}1$k random samples.

Note that in our current approach parameter and edge generation are completely independent. However, the types and counts of graph nodes give strong cues about which parameters and edges to generate. Our results show that this already generates realistic graphs. Conditioning parameter generation on edges is an interesting direction for future work, for example by describing the local graph neighborhood around a node with a graph neural network.

\begin{figure*}[t]
    \centering
    \includegraphics[width=\textwidth]{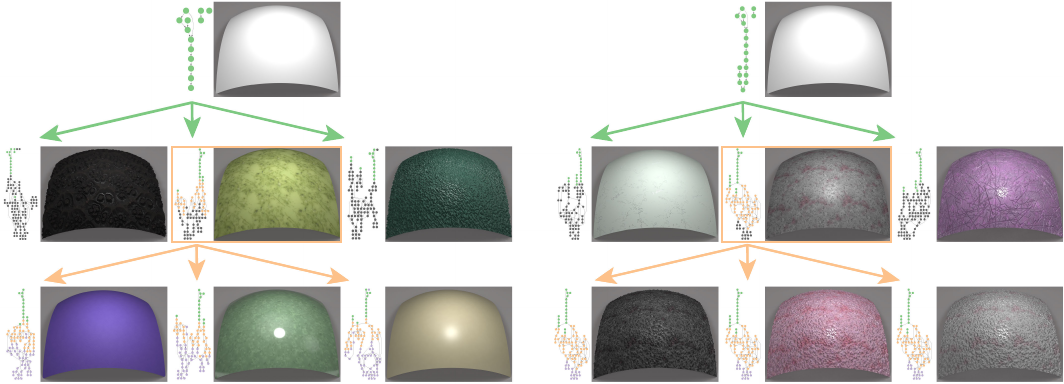}
    \caption{
    \textbf{Guided material authoring. }
    We show two examples of a novel guided-authoring workflow. Starting from a partial material graph (top row), we generate three completions of the graph using our method (middle row). Material graphs are shown next to their output material. Nodes are illustrated as colored circles, where the color indicates node provenance. Green nodes come from the initial partial graph. The user selects one of the suggested completions (orange border), selects a subset of the nodes they want to keep (orange), and ask for another completion round (bottom row). In the left example, the final round of completions re-generates a large part of the graph (purple nodes), resulting in a varied material appearance. In the right example, a smaller part of the graph is re-generated, resulting in similar material structure for all three completions, since in this example, the structure is determined by the nodes that were held fixed (orange).}
    \label{fig:autocomplete}
\end{figure*}

\subsubsection{Editing materials.}
Since the generated materials are procedural, we have full control over their appearance. Figure~\ref{fig:editing} shows examples of editing generated materials. We performed two edits on each of the four materials using Substance Designer~\cite{Substance}, each editing session taking less than $10$ minutes. Most of the time was spent on re-arranging the spatial layout of nodes (which is generated with a standard graph layout tool) in UI. The actual edits required at most $10$ slider changes. The edits add diverse changes to the materials, such as weathering effects, color changes, or structure changes, such as different brick sizes, an increased number of wooden planks, and changed wood grain structure.

\subsubsection{User Study.}
In addition to the editing examples, we evaluate the practical usability of our generated graphs compared to the baselines with a user study. We contacted two professional material artists to evaluate generated graphs in a blind user study. Each artist was shown 20 graphs generated by \name ($\pi_r$ node order) and 10 graphs from each baseline, and tasked to (i) rate the usability of graph on a scale from 0 to 5, with 5 being the best score, and (ii) give a short feedback on each graph. The artists could inspect and interact with each graph as long as they wanted in Substance Designer. We received a score of $1.82$ (40\% of scores $\ge3$) compared to 0.35 (10\% scores $\ge3$) for the Macro-nodes and 0.48 (0\% scores $\ge3$) for the  Pairwise baselines. This demonstrates the significant improvement provided by our method, as well as showing that there is still room for further research. Most prominent feedback for our method was that some materials are good and interesting, some are hard to understand, some produce oversaturated albedos, and some have unrealistic normals.

\subsection{Guided Material Authoring}
We can condition \name on a partial graph by using the set of existing nodes and edges as partial input sequences $s_{<i}$. This enables an interesting application where our method provides guidance to a user that is working on a partial graph. We can suggest multiple different completions of the graph, and can show the final result of each completion (similar to using autocomplete in a text editor).
The user can then select either all nodes of a preferred completion, obtaining a completed graph, or select a subset of the nodes in the completions, and continue exploring additional completions. If more nodes are selected from the completion, more properties of the material are held fixed in the following completions.

A few example completions are shown in Figure~\ref{fig:autocomplete}. Here we use a model trained with the node ordering $\pi_{rr}$; since users typically start with the front of the material graph, we need a front-to-back ordering. Each tree shown in the figure corresponds to two rounds of graph completions. We start with the graph at the root, and create three different completions~(second row). A preferred completion is then selected, together with a subset of nodes (orange), and a second round of completions is performed, shown in the bottom row. The node color indicates which of the tree levels each node comes from: green is the top (root) level, orange is the middle level, purple the bottom level. Gray nodes are not passed to the next level.

In the examples shown in the Figure~\ref{fig:autocomplete}, we randomly select the nodes at each level of the tree.
This approach can help novice users and can also be used by advanced users to explore material spaces.

%% file: sections/conclusion.tex
\section{Conclusion and Future Work}

We have presented \name---a transformer-based autoregressive model that can generate material graphs, consisting of operation nodes, node parameters, and edges that describe information transmission between nodes.
We quantitatively and qualitatively evaluated the quality and diversity of the generated materials. 
As an application, we presented a novel guided-authoring workflow where the user can progressively create complex material graphs by selecting from autocompletions proposed by our generator.
While \name is constrained by the diversity of the procedural graphs it is trained on, we expect it to improve as these datasets expand.

\name is, to the best of our knowledge, the first generative model for procedural material graphs. We believe it is an important step towards bringing the power of deep generative models to a critical part of 3D content creation pipelines. We also believe that our work opens up a number of important research directions that we would hope will be addressed in the course of future efforts. 

\paragraph{Projecting images to material graphs.}
In its present form, \name does not support matching a target material image by conditioning the generation on a target image.
Unlike GAN inversion in the context of images~\cite{abdal2019image2stylegan}, here the task is more complex due to the autoregressive generator setup, and the fact that generating an image requires evaluating the \emph{entire} graph.
An efficient scheme to backpropagate from images to graphs would enable many possibilities including image-conditioned graph generation and training graph generation supervised by only (large) image datasets instead of (much smaller) procedural material datasets.

A possible avenue is to train another network (i.e., an encoder network) to predict intermediate subgraphs of the target graph, and then use the proposed autocomplete framework, possibly in conjunction with a beam search, to iteratively search and explore the solution space. Once we obtain a good material graph structure, we can do a parameter refinement using a material-specific optimization setup~\cite{Shi:2020:Match}. A challenge here is to generate material graphs that preserve any mesostructure contained in the target images, possibly requiring a new material similarity metric. 

\paragraph{Semantic control handles.}
\name can be sampled to produce novel material graphs along with default parameters. However, these graphs often 
span 50+ nodes, making them complex to work with, without knowing the effect of the different node parameters. Artists, in contrast, often expose a small subset of parameter sliders, usually with semantic associations, to simplify editing the material graphs. In the future, it would be interesting to similarly expose a subset of the parameters, along with relevant value ranges, ideally with semantic attributes similar to latent spaces for portrait editing with StyleGAN~\cite{ganSpace:2020,styleFlow:20}.

\paragraph{Discovered macros.}
An interesting next step would be to discover commonly recurring subgraphs across multiple generated graphs. If successful, this would allow discovering \textit{metanodes}, which capture group of operations recurring across examples in the training set. This is similar to the problem of program synthesis where program macros are extracted by considering e-graphs of equivalent operations~\cite{dreamCoder:2020} in a bottom-up fashion. If successful, automatically discovered high-level metanodes can be reused for more abstracted material graph generation. 

\paragraph{Hierarchical output.}
At present, an end user has no guidance as to what the different parts of a generated graph represent. As a result, manipulating the graph remains tedious. In the future, we would like to produce hierarchically structured graphs. Similar to coarse-to-fine synthesis in the case of progressive GANs, a possible option is to have stages to first produce coarse (macro) structures, then add local properties (e.g., color, shininess), and finally spatially varying effects arising from weathering. An immediate challenge is to find training data for this hierarchy.

%% file: sections/appendix.tex
\appendix

\section{Semantic Validity Checks}
\label{app:semantic_validity}

\paragraph{Parameter generation.}
We use the semantic validity approach described in Section~\ref{sec:transformers} to constrain generated token values to represent valid node graphs. We do not perform semantic validity checks for the node generation step. For node parameter generation, we constrain the value of discrete parameters within their valid range. Similarly, we constrain the parameter indices in the sequence $S^\text{pk}$ to not exceed the number of number of parameters in the corresponding operator $o_\tau$.

\paragraph{Edge generation.}
We use the semantic validity approach to constrain edges to always go from output slots to input slots, i.e., the first element of an edge tuple has to refer to an output slot, the second element to an input slot. Additionally, we make sure edges do not form cycles by constraining edge end points to nodes that are not ancestors of the node at the edge start point.

\section{Conditional Transformer Generator}
\label{app:conditioning}
To condition a Transformer generator on a feature vector, we add three layers to each feed-forward block of the transformer (see Viswani et al.~\shortcite{Vaswani:2017:Transformers} for a detailed description of the Transformer blocks). The original block performs the following operations:
\begin{equation}
    \texttt{FFBlock}(x) \coloneqq x + \text{MLP}(\text{LN}(x)),
\end{equation}
where $x$ is the input feature vector, $\text{MLP}$ is a Multilayer Perceptron~\cite{hastie2009elements}, and $\text{LN}$ is Layer Normalization~\cite{ba2016layer}. We modify this block to take the condition as additional feature vector $c$:
\begin{equation}
    \texttt{CFFBlock}(x, c) \coloneqq \texttt{FFBlock}(x) + \text{MLP}(\text{LN}(c)),
\end{equation}

\section{Graph Statistics}
\label{app:graph_stats}
We compute the following graph statistics:
\begin{enumerate}
    \item The number of nodes.
    \item The number of disconnected components.
    \item The longest possible path in the graph.
\end{enumerate}
Additionally, several statistics that are computed per operator type, or per pair of operator types give more fine-grained information:
\begin{enumerate}
    \item The number of nodes for each operator type in a graph.
    \item The graph distance between each node of a given operator type and the closest output node.
    \item The number of connected input slots for each operator type.
    \item The graph distance between all pairs of operator types.
\end{enumerate}

\section{Architecture and Training Details}
\label{app:arch_details}

\paragraph{Architecture.}
All our transformer models use the GPT-2 architecture~\cite{radford2019language} with $2$ attention layers, $4$ attention heads in each layer, and $64$-dimensional hidden features. We use a maximum sequence length (not counting start/stop tokens) of $400$ nodes for the node generator $f^\text{n}$ and node encoder $g^\text{p}$, $512$ scalar parameter values for the parameter generator $f^\text{p}$, $700$ edges ($1400$ tokens) for the edge generator $f^\text{e}$, and $800$ slots for the slot encoder $g^\text{e}$. The number of different node types in our experiments is $1207$, this is also the vocabulary size of the main input sequence of the node generator, node encoder, and slot encoder.

\paragraph{Training.}
We train all models with the Adam optimizer~\cite{diederik2014adam} using a learning rate of $1\mathrm{e}{-4}$, a batch size of $64$ for the node and parameter generators, and a batch size of $16$ for the edge generator. For all three generators, we stop training once a minimum of the validation loss is reached, which happens in our experiments after $30$ epochs for the node generator, $11$ epochs for the parameter generator, and $10$ epochs for the edge generator. With one V100 GPU for each of the three models, training takes roughly 2-3 days.